# A Dynamic Equivalent Energy Storage Model of Natural Gas Networks for Joint Optimal Dispatch of Electricity-Gas Systems

Siyuan Wang, *Graduate Student Member, IEEE,* Wenchuan Wu, *Fellow, IEEE,*
Chenhui Lin, Binbin Chen

*Abstract*—The development of energy conversion techniques enhances the coupling between the gas network and power system. However, challenges remain in the joint optimal dispatch of electricity-gas systems. The dynamic model of the gas network, described by partial differential equations, is complex and computationally demanding for power system operators. Furthermore, information privacy concerns and limited accessibility to detailed gas network models by power system operators necessitate quantifying the equivalent energy storage capacity of gas networks. This paper proposes a multi-port energy storage model with time-varying capacity to represent the dynamic gas state transformation and operational constraints in a compact and intuitive form. The model can be easily integrated into the optimal dispatch problem of the power system. Test cases demonstrate that the proposed model ensures feasible control strategies and significantly reduces the computational burden while maintaining high accuracy in the joint optimal dispatch of electricity-gas systems. In contrast, the existing static equivalent model fails to capture the full flexibility of the gas network and may yield infeasible results.

## NOMENCLATURE

### A. Variables

| | |
|---|---|
| $\rho, v, \pi$ | Density, velocity and pressure of gas |
| $\pi_{i,t}, f_{i,t}$ | Gas pressure and mass flow rate at the $i$-th computation node at time $t$ |
| $s_t$ | Vector composed of the gas pressure and mass flow rate of all the uncontrollable computation nodes at time $t$ |
| $u_t$ | Vector composed of all the controllable gas pressure and injection mass flow rate of normal nodes at time $t$ |
| $\pi_{n,t}^{\text{NODE}}, f_{n,t}^{\text{INJ}}$ | Gas pressure and mass flow injection at normal node $n$ at time $t$ |
| $\Delta\pi_{k,t}^{\text{COMP}}$ | Boosted gas pressure provided by the $k$-th compressor at time $t$ |
| $f_{k,t}^{\text{GW}}, \pi_{k,t}^{\text{GW}}$ | Output mass flow rate and gas pressure of the $k$-th gas well at time $t$ |
| $f_{k,t}^{\text{GT}}, p_{k,t}^{\text{GT}}$ | Input gas mass flow rate and output power and of the $k$-th gas turbine at time $t$ |
| $f_{k,t}^{\text{P2G}}, p_{k,t}^{\text{P2G}}$ | Output gas mass flow rate and input power of the $k$-th power to gas unit at time $t$ |
| $\boldsymbol{p}_t^{\text{DEV}}$ | Vector composed of all the active power of GTs and P2Gs at time $t$ |
| $\boldsymbol{\pi}_t^{\text{CTRL}}$ | Vector of controllable pressure variables at time $t$, composed of the pressure of all GWs and boosted pressure provided by all the compressors |
| $\boldsymbol{f}_t^{\text{LD}}$ | Vector composed of the forecasted non-power gas load for all normal nodes |
| $\boldsymbol{z}_t^{\text{CTRL}}$ | Vector collects all the variables including power of GTs and P2Gs, and controllable gas pressure for the first $t$ time slots |
| $\boldsymbol{p}^{\text{GAS}}$ | Vector composed of the injection power from the gas network $p_t^{\text{GAS}}$ of all time slots |
| $\boldsymbol{P}^{\text{DEV}}$ | Matrix composed of all the active power of GTs and P2Gs of all time slots |

### B. Parameters

| | |
|---|---|
| $u$ | Speed of sound |
| $\lambda$ | Friction coefficient of gas pipeline |
| $v_b$ | Base value of gas velocity of gas pipeline |
| $A, D, \lambda, \alpha$ | Cross-sectional area, diameter, friction coefficient and inclination of gas pipeline |
| $q_L$ | Lower heat value of natural gas |
| $\eta_k^s$ | Energy conversion efficiency of the $k$-th unit $s$, $s = \{\text{GT,P2G}\}$. |
| $\underline{f}_k^s, \overline{f}_k^s$ | Lower and upper bounds of gas mass flow rate injection of the $k$-th unit $s$, $s = \{\text{GT,P2G,GW}\}$. |
| $\underline{\pi}_k^s, \overline{\pi}_k^s$ | Lower and upper bounds of gas pressure of the $k$-th unit $s$, $s = \{\text{GT,P2G,GW}\}$. |
| $\underline{p}_k^s, \overline{p}_k^s$ | Lower and upper bounds of active power of the $k$-th unit $s$, $s = \{\text{GT,P2G}\}$. |
| $\Delta\underline{\pi}_k, \Delta\overline{\pi}_k$ | Lower and upper bounds of the $k$-th compressor's pressure booster limits |
| $\underline{\pi}_l, \overline{\pi}_l$ | Lower and upper bounds of the $l$-th pipeline's gas pressure |
| $\boldsymbol{W}_t, \boldsymbol{w}_t$ | Parameters by collecting the operational constraints of gas network for the first $t$ time slots |
| $\boldsymbol{D}_t$ | Constant matrix used to select the elements in |

This work was supported by the Science-Technique Plan Project of Jiangsu Province (No. SBE2022110130) and the National Science Foundation of China (No. U2066601).

S. Wang, W. Wu (Corresponding author, wuwench@tsinghua.edu.cn), C. Lin and B. Chen are with the State Key Laboratory of Power Systems, Department of Electrical Engineering, Tsinghua University, Beijing 100084, China.

| | |
|---|---|
| | $p_t^{\text{DEV}}$ from the vector $z_t^{\text{CTRL}}$ |
| $E_t$ | Parameter of the high-dimensional quadrant ellipsoid region $\mathcal{E}_t$ |
| $T$ | Total number of all time slots |
| $\underline{p}_t^{\text{GAS}}$, $\overline{p}_t^{\text{GAS}}$ | Lower and upper power bounds of the equivalent storage model of the gas network at time $t$ |
| $\underline{e}_t^{\text{GAS}}$, $\overline{e}_t^{\text{GAS}}$ | Lower and upper energy bounds of the equivalent storage model of the gas network at time $t$ |
| $A^{\text{GAS}}$, $b^{\text{GAS}}$ | Parameters of the equivalent energy storage model with time-varying capability |

*C. Sets and Functions*

| | |
|---|---|
| $\mathcal{L}$ | Index set of pipelines |
| $\mathcal{N}$ | Index set of normal nodes |
| $\mathcal{S}$ | Index set of all computation nodes |
| $\mathcal{S}_n$, $\mathcal{S}_l$ | Index set of variables at normal node $n$ and pipeline $l$ |
| $\mathcal{S}_n^+$, $\mathcal{S}_n^-$ | Index sets of variables that flow into and out of normal node $n$. |
| $\mathcal{S}_k^+$, $\mathcal{S}_k^-$ | Index of computation node flowing into and out of the compressor $k$ |
| $\mathcal{D}^s$ | Index set of all the units $s$ in the gas network, $s=\{\text{GT,P2G,GW,COMP,GEN}\}$ |
| $\mathcal{D}_n^s$ | Index set of all the units $s$ connected to the normal node $n$, $s=\{\text{GT,P2G,GW}\}$ |
| $\mathcal{T}$ | Index set of all time slots. |
| $\mathcal{E}_t$ | Power coupling region among different energy exchange interfaces of time $t$ |
| $\Omega^{\text{GAS}}$ | Controllability feasible region of gas network |
| $\mathcal{B}^{\text{GAS}}$ | Energy storage flexibility region of gas network |
| $\mathcal{F}^{\text{GAS}}$ | Integrated multi-port energy storage model with time-varying capacity of gas network |
| $(\bullet)_i$ | The $i$-th row of a matrix or the $i$-th element of a vector |
| $\text{card}(\bullet)$ | Cardinality of a set. |

## I. INTRODUCTION

*A. Motivation*

SINCE energy conversion techniques related to the integrated energy system were well developed [1], [2], which makes the gas network and power system coupled more tightly [3]. Exploiting the inherent flexibility of gas networks can offer valuable opportunities to the power system, enabling it to provide energy storage services [4], [5] and maintain power balance.

However, for the power system, the challenges remain in leveraging the flexibility of gas networks. Firstly, the gas network and the power grid are generally operated separately [6], each associated with separate entities of interest, thereby raising concerns about information confidentiality [7] between these two systems. Consequently, the power system operator is unable to directly utilize the intricate model of gas networks for cohesive coordination [8]. Furthermore, the gas network employs partial differential equations to describe its dynamic processes [9], imposing a substantial computational burden on the joint dispatch of the electricity-gas system. Additionally, the gradual and inert nature of the gas grid system poses difficulties in quantifying the energy storage capacity essential for power system operators.

In this paper, we propose a method to evaluate the equivalent energy storage model of gas networks. The slow dynamic process and operational constraints of the gas network are transformed into coupling power constraints and energy constraints among the active power of converters in the interface between the power system and the gas networks, such as GTs and P2Gs. These constraints constitute the equivalent energy storage model of the gas network, which can be easily incorporated into the optimal dispatch model of the power system.

*B. Literature review*

The flexibility or equivalent model of gas network has been studied in serval works. [10] and [11] used the energy hub as the energy conversion port and estimated steady-state security region. The nodal operating envelope is used in [12] to represent the aggregated flexibility of the integrated energy units, taking into account multiple P2Gs and the operational constraints of gas networks. The work in [13] presented a comprehensive overview for the flexibility of distributed multi-energy systems and provided a general method to aggregate their flexibility. The inertia of gas and thermal is illustrated in [14] based on the corresponding dynamic model. The zonal linepack is introduced in [15] to quantify the gas flexibility and applied in the optimal power flow model in the integrated gas and electrical system. In [16], an outer approximation with equality relaxation method is presented to speed up the integrated operation of multi-energy systems. A feasible region composed of a large number of predefined constraints of the natural gas-fired units is proposed in [17], and this region are then used to replace the natural gas network in the electricity-gas co-optimization problem to simplify the calculation.

The current solutions have facilitated the effective evaluation of the flexibility model of gas networks in certain aspects. However, the primary drawback of existing methods lies in their reliance on static models, ignoring the time-coupling constraints inherent in the gas network. Temporal coupling constraints pertain to the influence of preceding states on subsequent states of the gas system, ensuring the validity of the gas network's state space equation. Consequently, the two adjacent states cannot be considered independent entities. By considering temporal coupling constraints, we can achieve a more accurate depiction of the transient process of the gas network, thereby avoiding impractical solutions arising from sudden changes in gas state and reducing errors in the optimal coordination of electricity-gas systems. Moreover, in order to fully account for the dynamic nature of the gas network, the current models that consider dynamic processes can only employ detailed network models, leading to heavy computation burden in handling the dynamic gas state. It is necessary to substitute a simplified equivalent energy storage model for the detailed network model, thereby mitigating the substantial computational burden associated with computing the dynamic processes of the gas network.

*C. Contribution and paper organization*

This paper presents a novel solution to evaluate the equivalent energy storage model of gas networks. Since the coupling interface between the power system and the gas network may involve several GTs and P2Gs, the gas network can be equivalent to a multi-port energy storage model with time-varying capacity. The regulation capability of GTs and P2Gs can be captured by a flexibility region, which is formulated by the state-space model of the natural gas network and its operational constraints. Then, the state-space model and operational constraints are represented in a reduced and more intuitive form.

To the best knowledge of the authors, the main contributions of our work are as follows:

(1) A multi-port energy storage model with time-varying capacity is proposed to quantify the flexibility of the gas network, which implicitly incorporates all operational constraints. Therefore, it can be easily incorporated into the optimal dispatch problem of the power system to preserve the information privacy and reduce the computational burden.

(2) The inscribed high-dimensional quadrant ellipsoid algorithm is adopted to describe coupling relationships among the active power of energy converters, such as GTs and P2Gs. It offers significant advantages in terms of both approximation accuracy and computational complexity.

(3) The high-dimensional polyhedron projection and bounds shrinking algorithm is developed to calculate the parameters of the equivalent energy storage model. The flexibility of gas networks can be fully exploited and the feasibility of the resulted control strategies can be guaranteed.

The remainder of this paper is organized as follows. Gas network models with gas sources and units are formulated in Section II. The joint dispatch model of the gas and power network is introduced in Section III. Section IV delineates the detailed process for calculating the equivalent energy storage model. Numerical tests are presented in Section V and conclusions are made in Section VI.

## II. FORMULATION OF THE GAS NETWORK AND UNITS

The gas states along a pipeline obey the partial differential equations [18] as follows (supposing $v(x,t) > 0$):

$$\frac{\partial \rho(x,t)}{\partial t} + \frac{\partial \rho(x,t)v(x,t)}{\partial x} = 0 \quad (1)$$

$$\frac{\partial \rho(x,t)v(x,t)}{\partial t} + \frac{\partial \rho(x,t)v^2(x,t)}{\partial x} + \frac{\partial \pi(x,t)}{\partial x} + \frac{\lambda \rho(x,t)v^2(x,t)}{2D} + \rho(x,t)g \sin\alpha = 0 \quad (2)$$

where (1) represents the principle of mass conservation in a natural gas pipeline, signifying that the net outflow rate of mass from an infinitesimal fluid volume is equivalent to the rate of mass decrement within the differential volume. (2) represents the laws of momentum. The five terms correspond to the natural gas inertia, convective gas flow, dynamic gas flow pressure, hydraulic friction force, and force of gravity, respectively. $\pi(x,t)$, $\rho(x,t)$ and $v(x,t)$ denote the gas pressure, density and velocity at position $x$ at time $t$, respectively; $D$, $\lambda$ and $\alpha$ denote the cross-sectional diameter, friction coefficient and inclination of the gas pipeline, respectively.

Utilizing the first-order Taylor expansion, the quadratic term can be expanded at the base value of gas velocity $v_b$ [19], [20]. The linearization error resulting from the Taylor expansion is considerably smaller than 1%, thus it can be disregarded.

$$v^2(x,t) \approx 2v(x,t)v_b - v_b^2 \quad (3)$$

Besides, the second convective term in (2) can be approximately considered as 0 [21], and the inclination angle $\alpha$ is 0, too. The gas pressure and mass flow rate can be expressed by the state variables as follows:

$$\pi(x,t) = u^2 \rho(x,t) \quad (4)$$

$$f(x,t) = A\rho(x,t)v(x,t) \quad (5)$$

where $A$ denotes the cross-sectional area of the gas pipeline.

By substituting (3)-(5) into (1)-(2), the partial differential equations of gas states can be expressed as follows:

$$A\frac{\partial \pi(x,t)}{\partial t} + u^2 \frac{\partial f(x,t)}{\partial x} = 0 \quad (6)$$

$$\frac{1}{A}\frac{\partial f(x,t)}{\partial t} + \frac{\partial \pi(x,t)}{\partial x} + \frac{\lambda v_b}{AD}f(x,t) - \frac{\lambda v_b^2}{2u^2 D}\pi(x,t) = 0 \quad (7)$$

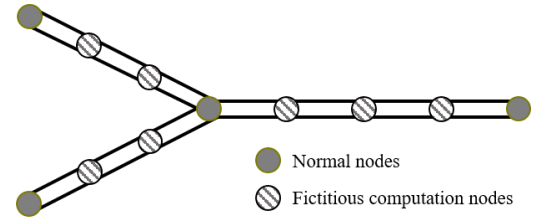

Fig. 1. Computation nodes of gas pipeline network.

For the simplification of numerical calculation, the partial differential equations can be discretized in space and time. As shown in Fig. 1, there are two kinds of computation nodes, normal nodes and fictitious computation nodes [22]. The normal nodes are terminal nodes of each gas pipelines, and the fictitious computation nodes are obtained by discrete segmentation of the gas pipelines. Then, the state-space model (6)-(7) can be described as the discrete difference equations as follows:

$$A\frac{\pi_{i,t} - \pi_{i,t-1}}{\Delta t} + u^2 \frac{f_{i,t} - f_{i-1,t}}{\Delta x} = 0 \quad (8)$$

$$\frac{f_{i,t} - f_{i,t-1}}{A\Delta t} + \frac{\pi_{i,t} - \pi_{i-1,t}}{\Delta x} + \frac{\lambda v_b}{AD}f_{i,t} - \frac{\lambda v_b^2}{2u^2 D}\pi_{i,t} = 0 \quad (9)$$

Besides, there are topology constraints of gas pipeline networks as follows:

$$\pi_{i,t} = \pi_{n,t}^{\text{NODE}}, i \in \mathcal{S}_n \tag{10}$$

$$\sum_{i \in \mathcal{S}_n^+} f_{i,t} + f_{n,t}^{\text{INJ}} = \sum_{j \in \mathcal{S}_n^-} f_{j,t} \tag{11}$$

where $\pi_{n,t}^{\text{NODE}}$ denotes the gas pressure of normal node $n$ at time $t$; $f_{n,t}^{\text{INJ}}$ denotes the injection mass flow rate of normal node $n$ at time $t$; $\mathcal{S}_n$ denotes the index set of variables at normal node $n$; $\mathcal{S}_n^+$ and $\mathcal{S}_n^-$ denote the index sets of variables that flow into and out of normal node $n$.

The discrete difference equations (8)-(9) and topology constraints (10)-(11) can be transformed into the compact matrix form [22] as follows:

$$\boldsymbol{s}_t = \boldsymbol{J}\boldsymbol{s}_{t-1} + \boldsymbol{H}\boldsymbol{u}_{t-1} \tag{12}$$

where $\boldsymbol{s}_t$ is a vector composed of the gas pressure and mass flow rate of all the uncontrollable computation nodes at time $t$. $\boldsymbol{u}_t$ is a vector composed of all the controllable gas pressure and injection mass flow rate of normal nodes at time $t$. $\boldsymbol{J}$ and $\boldsymbol{H}$ are constant matrices obtained based on (8)-(11).

For the consideration of security, the gas pressures of gas pipelines also have upper and lower bounds. For any $i \in \mathcal{S}_l$

$$\underline{\pi}_l \leq \pi_{i,t} \leq \overline{\pi}_l \tag{13}$$

The gas injection at nodes can be described as the net injection of all the units. For any $n \in \mathcal{N}$

$$f_{n,t}^{\text{INJ}} = \sum_{k \in \mathcal{D}_n^{\text{GW}}} f_{k,t}^{\text{GW}} + \sum_{k \in \mathcal{D}_n^{\text{P2G}}} f_{k,t}^{\text{P2G}} - \sum_{k \in \mathcal{D}_n^{\text{GT}}} f_{k,t}^{\text{GT}} - f_{n,t}^{\text{LD}} \tag{14}$$

where $\mathcal{D}_n^{\text{GW}}$, $\mathcal{D}_n^{\text{P2G}}$, and $\mathcal{D}_n^{\text{GT}}$ denote the index sets of gas sources, P2Gs, and GTs that directly linked to node $n$, respectively. $f_{n,t}^{\text{LD}}$ denotes the mass flow rate of forecasted non-power gas load at node $n$ at time $t$.

As the gas source, the gas wells (GWs) provide natural gas for the gas network with bounded mass flow rate and gas pressure. For any $k \in \mathcal{D}^{\text{GW}}$

$$\underline{f}_k^{\text{GW}} \leq f_{k,t}^{\text{GW}} \leq \overline{f}_k^{\text{GW}} \tag{15a}$$

$$\underline{\pi}_k^{\text{GW}} \leq \pi_{k,t}^{\text{GW}} \leq \overline{\pi}_k^{\text{GW}} \tag{15b}$$

The compressors are used to boot the gas pressure in the gas pipeline, which have the maximum and minimum pressure booster limits. For any $k \in \mathcal{D}^{\text{COMP}}$

$$\Delta\underline{\pi}_k \leq \Delta\pi_{k,t} \leq \Delta\overline{\pi}_k \tag{16}$$

The gas turbines (GTs) consume natural gas and generate electricity with the efficiency $\eta_k^{\text{GT}}$. For any $k \in \mathcal{D}^{\text{GT}}$

$$p_{k,t}^{\text{GT}} = \eta_k^{\text{GT}} f_{k,t}^{\text{GT}} q_L \tag{17a}$$

$$\underline{p}_k^{\text{GT}} \leq p_{k,t}^{\text{GT}} \leq \overline{p}_k^{\text{GT}} \tag{17b}$$

$$\underline{f}_k^{\text{GT}} \leq f_{k,t}^{\text{GT}} \leq \overline{f}_k^{\text{GT}} \tag{17c}$$

where $q_L$ denotes the low heat value of natural gas.

The power-to-gas devices (P2Gs) consume electricity and generate natural gas with the efficiency $\eta_k^{\text{P2G}}$. For any $k \in \mathcal{D}^{\text{P2G}}$

$$f_{k,t}^{\text{P2G}} = \eta_k^{\text{P2G}} p_{k,t}^{\text{P2G}} / q_L \tag{18a}$$

$$\underline{p}_k^{\text{P2G}} \leq p_{k,t}^{\text{P2G}} \leq \overline{p}_k^{\text{P2G}} \tag{18b}$$

$$\underline{f}_k^{\text{P2G}} \leq f_{k,t}^{\text{P2G}} \leq \overline{f}_k^{\text{P2G}} \tag{18c}$$

Based on the definition of control vector $\boldsymbol{u}_t$ in (12), the model of GTs and P2Gs in (17a) and (18a), and the definition of gas injection $f_{n,t}^{\text{INJ}}$ in (14), the state vector $\boldsymbol{s}_t$ can be reformulated as the affine form of state vector $\boldsymbol{s}_{t-1}$, power of GTs and P2Gs $\boldsymbol{p}_t^{\text{DEV}}$, controllable gas pressure $\boldsymbol{\pi}_t^{\text{CTRL}}$ (including the gas pressure of GWs and boosted gas pressure provided by compressors), and forecasted gas consumption of non-controllable gas loads $\boldsymbol{f}_t^{\text{LD}}$.

$$\boldsymbol{s}_t = \boldsymbol{B}_1\boldsymbol{s}_{t-1} + \boldsymbol{B}_2\boldsymbol{p}_t^{\text{DEV}} + \boldsymbol{B}_3\boldsymbol{\pi}_t^{\text{CTRL}} + \boldsymbol{B}_4\boldsymbol{f}_t^{\text{LD}} \tag{19}$$

where $\boldsymbol{B}_1 \sim \boldsymbol{B}_4$ are constant matrices; $\boldsymbol{f}_t^{\text{LD}}$ is a vector composed of $f_{n,t}^{\text{LD}}$ for all $n \in \mathcal{N}$; $\boldsymbol{p}_t^{\text{DEV}}$ denotes a vector composed of all the active power of GTs and P2Gs at time $t$, that is

$$\boldsymbol{p}_t^{\text{DEV}} := \left[\cdots, p_{i,t}^{\text{GT}}, \cdots, p_{j,t}^{\text{P2G}}, \cdots\right]^\top, i \in \mathcal{D}^{\text{GT}}, j \in \mathcal{D}^{\text{P2G}} \tag{20}$$

$n_{dev} := \text{card}(\mathcal{D}^{\text{GT}}) + \text{card}(\mathcal{D}^{\text{P2G}})$; $\boldsymbol{\pi}_t^{\text{CTRL}}$ denotes the vector of controllable pressure variables at time $t$, composed of the pressure of all GWs and boosted pressure provided by all the compressors, that is

$$\boldsymbol{\pi}_t^{\text{CTRL}} := \left[\cdots, \pi_{k,t}^{\text{GW}}, \cdots, \Delta\pi_{c,t}^{\text{COMP}}, \cdots\right]^\top \tag{21}$$
$$k \in \mathcal{D}^{\text{GW}}, c \in \mathcal{D}^{\text{COMP}}$$

Define the vector $\boldsymbol{z}_t^{\text{CTRL}}$ that collects all the variables including active power of GTs and P2Gs, and controllable gas pressure for the first $t$ time slots, that is

$$\boldsymbol{z}_t^{\text{CTRL}} := \left[\cdots, (\boldsymbol{p}_i^{\text{DEV}})^\top, (\boldsymbol{\pi}_i^{\text{CTRL}})^\top, \cdots\right]^\top, i = 1, 2, 3, \cdots, t \tag{22}$$

To simplify the expression, all the operational constraints in (12)-(19) can be reformulated for the first $t$ time slots as a matrix compact form as (23)-(24). The constraints (15b), (16), (17b) and (18b) related to the control variables correspond directly to the parameters in the matrix $\boldsymbol{W}_t$. The constraints (13) and (15a) related to the gas network state can be expressed as the linear form of control variables according to the state space equation (19).

$$W_t z_t^{CTRL} \leq w_t \quad (23)$$

$$p_t^{DEV} = D_t z_t^{CTRL} \quad (24)$$

where the matrix $W_t$ and vector $w_t$ are constant by collecting the constraints (12)-(19) for the first $t$ time slots; $D_t$ is a constant matrix used to select the elements of $p_t^{DEV}$ from the vector $z_t^{CTRL}$ according to its definition.

## III. JOINT DISPATCH MODEL OF THE GAS AND POWER NETWORK

The joint dispatch model of the gas and power network can be expressed as follows:

$$\min \sum_{t \in \mathcal{T}} \left( \sum_{i \in \mathcal{D}^{GEN}} C_i^{GEN} + \sum_{j \in \mathcal{D}^{GT}} C_j^{GT} \right) \cdot \Delta t \quad (25a)$$

$$s.t. \quad \underline{p}_k^{GEN} \leq p_{k,t}^{GEN} \leq \overline{p}_k^{GEN} \quad (25b)$$

$$-r_k^{GEN} \Delta t \leq p_{k,t}^{GEN} - p_{k,t-1}^{GEN} \leq r_k^{GEN} \Delta t \quad (25c)$$

$$N_t \begin{bmatrix} p_t^{DEV} \\ p_t^{GEN} \end{bmatrix} \leq n_t \quad (25d)$$

$$W_T z_T^{CTRL} \leq w_T \quad (25e)$$

$$p_t^{DEV} = D_t z_t^{CTRL} \quad (25f)$$

where (25a) is the objective function of optimal joint dispatch to minimize the total cost of fuel consumption; (25b) and (25c) denote the active power constraints and ramp constraints of generators. (25d) is a concise constraint that implies the linear flow constraints of the power network, including flow constraints for branches and voltage constraints for buses in the power network. (25e) and (25f) are the constraints of gas network. $C_i^{GEN}$ denotes the fuel cost of the $i$-th coal-fired power unit, as shown in (26a), $a_i^{GEN}$, $b_i^{GEN}$ and $c_i^{GEN}$ as the cost parameters; $C_j^{GT}$ denotes the gas consumption cost of $j$-th GT unit, as shown in (26b); $\lambda_j^{GT}$ as the cost of natural gas. The generation cost of renewable generators are considered as 0.

$$C_i^{GEN} = a_i^{GEN} \left( p_{i,t}^{GEN} \right)^2 + b_i^{GEN} p_{i,t}^{GEN} + c_i^{GEN} \quad (26a)$$

$$C_j^{GT} = \lambda_j^{GT} f_{j,t}^{GT} \quad (26b)$$

In the above joint dispatch problem, the foremost computational burden stems from the gas network state constraint (25e). This arduous task arises due to the intricate state equation depicted in (19), necessitating a multitude of auxiliary variables to represent the states of fictitious computation nodes. The calculation of each state variable needs to be performed point by point. Therefore, it becomes imperative to establish a simplified equivalent expression that encapsulates the gas network's flexibility, denoting it as $\mathcal{F}^{GAS}$. By doing so, the computation process in (25) can be significantly expedited. Consequently, the substitution of (25e) and (25f) is warranted, and the ensuing simplified expressions are as follows:

$$P^{DEV} \in \mathcal{F}^{GAS} \quad (27)$$

To establish the gas network constraints (25e), comprehensive information regarding the gas network and its devices is indispensable. Therefore, it is imperative for the gas network operator to transmit the data of gas network models to the power grid for dispatch purposes. Conversely, (27) presents a condensed model of flexibility. When participating in the joint dispatch, the gas network only needs to transmit the compressed flexibility model, thereby effectively protecting the data privacy of the gas network.

## IV. EQUIVALENT ENERGY STORAGE MODEL FOR THE GAS NETWORK

### A. Overview

Since the gas network has large capacity for storing the natural gas, it has the potential of energy storage. In addition, the gas network system converts energy with the power grid through GTs and P2Gs, corresponding to the discharging and charging process of energy storage devices, respectively. Therefore, the gas network model can be reformulated as an equivalent multi-port energy storage model with time-varying capacity.

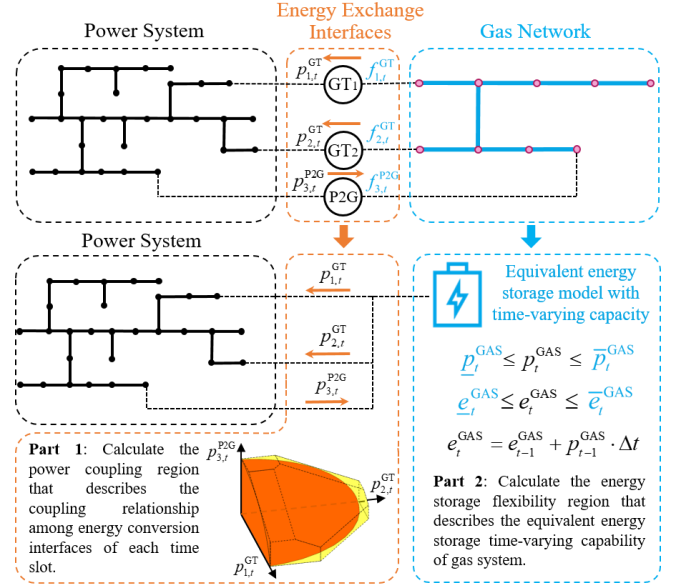

Fig. 2. Schematic of calculation process of the equivalent energy storage model.

As shown in Fig. 2, the process of evaluating the equivalent model of gas network is divided into two parts: the power coupling region of the energy conversion devices (GTs and P2Gs), and the energy storage flexibility region. In the first part, the power coupling region is calculated to describe the coupling relationship among the active power of the energy conversion devices of each time slot. In the second part, the energy storage flexibility region is formulated to describe the equivalent energy storage time-varying capability of the gas network.

## B. Calculate the power coupling region of converters

This subsection introduces the method to calculate power coupling region among different energy conversion devices of each time slot. A high-dimensional quadrant ellipsoid is used to approximate this region.

For each time slot, we try to find a region in the space of vector $p_t^{\text{DEV}}$ as large as possible, denoted as $\mathcal{E}_t$. All the points in this region should be feasible. That is to say, for any operation point $p_t^{\text{DEV}}$ inside $\mathcal{E}_t$, there is a corresponding feasible control solution $z_t^{\text{CTRL}*}$ that can realize the operation point $p_t^{\text{DEV}}$ while meeting all the operational constraints of the gas network. Its corresponding mathematical expression is: for $\forall p_t^{\text{DEV}} \in \mathcal{E}_t$, $\exists z_t^{\text{CTRL}*}$, that makes (23) and (24) hold.

There are several existing approaches for obtaining the inscribed high-dimensional region, such as employing the inscribed hypercube [23], inscribed ellipsoid [24], and vertex enumeration [25] methods. However, both the inscribed hypercube and inscribed ellipsoid methods are excessively conservative, leading to significant approximation errors. On the other hand, the vertex enumeration method is only suitable for computing lower-dimensional inscribed models but imposes a heavy computational burden in this context. We use a high-dimensional quadrant ellipsoid to approximate the feasible region of $p_t^{\text{DEV}}$. This method offers significant advantages in terms of both approximation accuracy and computational complexity.

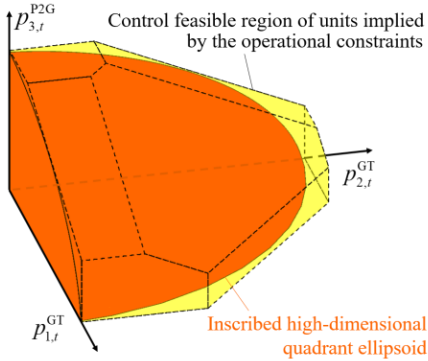

Fig. 3. Schematic of the inscribed high-dimensional quadrant ellipsoid.

The schematic is shown in in Fig. 3. Due to the operational constraints of the gas network, there may be a large number of potential coupling relationships among the active power of the energy conversion devices at time $t$. It is impossible to list all these constraints one by one explicitly, because the number of constraints increases exponentially with the number of devices. An alternative method is to approximate the exact expression of $\mathcal{E}_t$ with an inscribed high-dimensional quadrant ellipsoid. The ellipsoid is strictly inscribed in the original control feasible region, so that the feasibility of the resulted control strategies can be guaranteed.

This high-dimensional quadrant ellipsoid can be expressed as follows:

$$\mathcal{E}_t := \left\{ p_t^{\text{DEV}} \; \middle| \; \left\| E_t^{-1} \left( p_t^{\text{DEV}} - c \right) \right\| \leq 1, \; p_t^{\text{DEV}} \geq c \right\} \quad (28)$$

where $E_t$ is a positive semidefinite matrix that determined the shape of the ellipsoid; $c$ is a vector that determined the center of the ellipsoid, which composed of the minimum active power of each GTs ($\underline{p}_k^{\text{GT}}$) and P2Gs ($\underline{p}_k^{\text{P2G}}$).

Since the minimum active power constraints of each GTs and P2Gs are already reflected in the definition of the ellipsoid in (28), these constraints can be removed from the original constraints (23) and the remaining constraints are rewritten as:

$$\tilde{W}_t z_t^{\text{CTRL}} \leq \tilde{w}_t \quad (29)$$

To calculate the parameter $E_t$ of maximum volume ellipsoid $\mathcal{E}_t$ that inscribed in the original control feasible region, it can be transformed into the following semidefinite programming problem [24] as follows:

$$\max_{E_t, F, \gamma, \alpha} \; \log \det E_t \quad (30a)$$

$$s.t. \quad E_t \succeq 0 \quad (30b)$$

For $i = 1, 2, \cdots, n_w$

$$\left\| (W^{(1)})_i \tilde{D}^{-1} E_t + (W^{(2)})_i F \right\| \leq \alpha_i \quad (30c)$$

$$\alpha_i + (W^{(1)})_i \tilde{D}^{-1} c + (W^{(2)})_i \gamma - (w_t)_i \leq 0 \quad (30d)$$

where $F \in \mathbb{R}^{n_{\text{null}} \times n_{\text{dev}}}$, $\gamma \in \mathbb{R}^{n_{\text{null}}}$, and $\alpha \in \mathbb{R}^{n_w}$ are auxiliary decision variables; $n_{dev}$ and $n_w$ denote the length of vectors $p_t^{\text{DEV}}$ and $\tilde{w}_t$, respectively; $n_{\text{null}} := n_{\text{ctrl}} - n_{\text{dev}}$; $n_{\text{ctrl}}$ denotes the length of vector $z_t^{\text{CTRL}}$; Define the matrix $D^{(1)} \in \mathbb{R}^{n_{\text{ctrl}} \times n_{\text{dev}}}$ composed of the vectors that span the orthogonal bases of $D_t$, and matrix $D^{(2)} \in \mathbb{R}^{n_{\text{ctrl}} \times n_{\text{null}}}$ composed of the vectors that span the null-space of $D_t$; $W^{(1)} := \tilde{W}_t D^{(1)} \in \mathbb{R}^{n_w \times n_{\text{dev}}}$, $W^{(2)} := \tilde{W}_t D^{(2)} \in \mathbb{R}^{n_w \times n_{\text{null}}}$, $\tilde{D} := D_t D^{(1)} \in \mathbb{R}^{n_{\text{dev}} \times n_{\text{dev}}}$. The subscript $(\cdot)_i$ denotes the $i$-th row of a matrix or the $i$-th element of a vector. The detailed explanation of (30) can refer to [24].

## C. Calculate the energy storage flexibility region

From the perspective of the power system, the gas network can be equivalent to a multi-port energy storage model with time-varying capacity. This subsection introduces the method to calculate the parameters of time-varying capacity of all time slots.

Based on the constraints in (23), all the operational constraints of all the $T$ time slots can be expressed as follows:

$$W_T z_T^{\text{CTRL}} \leq w_T \quad (31)$$

where $T := \text{card}(\mathcal{T})$ denotes total number of all time slots.

Define the injection power from the gas network to the power system as the sum of all GTs' generation power minus all P2Gs' consumption power, that is

$$p_t^{\text{GAS}} := \sum_{i \in \mathcal{D}^{\text{GT}}} p_{i,t}^{\text{GT}} - \sum_{j \in \mathcal{D}^{\text{P2G}}} p_{j,t}^{\text{P2G}} = \boldsymbol{\eta}^\top \boldsymbol{p}_t^{\text{DEV}} \quad (32)$$

where $\boldsymbol{\eta}$ is a constant vector composed of the constant coefficients of each item in $\boldsymbol{p}_t^{\text{DEV}}$.

According to (22) and (32), the injection power from the gas network of all time slots $\boldsymbol{p}^{\text{GAS}}$ can be expressed in the linear form of control vector $\boldsymbol{z}_T^{\text{CTRL}}$ as follows:

$$\boldsymbol{p}^{\text{GAS}} = \boldsymbol{L} \boldsymbol{z}_T^{\text{CTRL}} \quad (33)$$

where $\boldsymbol{L}$ is a constant matrix and can be calculated based on (22) and (32).

To simplify the expression, the control feasible region that make (31) and (33) hold is denoted as $\Omega^{\text{GAS}}$.

$$\Omega^{\text{GAS}} := \left\{ (\boldsymbol{z}_T^{\text{CTRL}}, \boldsymbol{p}^{\text{GAS}}) \,\middle|\, \begin{matrix} \boldsymbol{W}_T \boldsymbol{z}_T^{\text{CTRL}} \leq \boldsymbol{w}_T \\ \boldsymbol{p}^{\text{GAS}} = \boldsymbol{L} \boldsymbol{z}_T^{\text{CTRL}} \end{matrix} \right\} \quad (34)$$

Similar to the work in Part 1, we try to find a region in the space of vector $\boldsymbol{p}^{\text{GAS}}$ as large as possible, denoted as $\mathcal{B}^{\text{GAS}}$ in this part. All the points in this region should guarantee the control strategies feasible. That is to say, for any operation point $\boldsymbol{p}^{\text{GAS}}$ inside $\mathcal{B}^{\text{GAS}}$, there is a corresponding feasible control solution $\boldsymbol{z}_T^{\text{CTRL}*}$ that can realize the operation point $\boldsymbol{p}^{\text{GAS}}$, while meeting the operational constraints of gas network in (31) and (33). That is, for $\forall \boldsymbol{p}^{\text{GAS}} \in \mathcal{B}^{\text{GAS}}$, $\exists \boldsymbol{z}_T^{\text{CTRL}*}$ such that $(\boldsymbol{z}_T^{\text{CTRL}*}, \boldsymbol{p}^{\text{GAS}}) \in \Omega^{\text{GAS}}$.

Since the gas network has the physical interpretation of energy storage, the equivalent model of the gas network can be approximated by the following energy storage form:

$$\underline{p}_t^{\text{GAS}} \leq p_t^{\text{GAS}} \leq \overline{p}_t^{\text{GAS}} \quad (35a)$$

$$\underline{e}_t^{\text{GAS}} \leq e_t^{\text{GAS}} \leq \overline{e}_t^{\text{GAS}} \quad (35b)$$

where

$$e_t^{\text{GAS}} = e_{t-1}^{\text{GAS}} - p_{t-1}^{\text{GAS}} \cdot \Delta t \quad (35c)$$

The equivalent energy storage device's parameters are determined collectively by the gas pipeline network parameters and each device within the gas system. Typically, a gas system equipped with higher-powered GTs and P2Gs devices correlates with a greater charging and discharging capacity of the equivalent energy storage. Furthermore, a gas network boasting a larger gas storage capacity is associated with a larger capacity of the equivalent energy storage.

The equivalent energy storage model (35) can be further simplified as the following matrix compact form:

$$\mathcal{B}^{\text{GAS}} := \left\{ \boldsymbol{p}^{\text{GAS}} \,|\, \boldsymbol{A}^{\text{GAS}} \boldsymbol{p}^{\text{GAS}} \leq \boldsymbol{b}^{\text{GAS}} \right\} \quad (36)$$

where $\boldsymbol{p}^{\text{GAS}}$ is a vector collecting the $p_t^{\text{GAS}}$ of all the time steps; $\boldsymbol{A}^{\text{GAS}}$ is a constant matrix denoted as the coefficients of $p_t^{\text{GAS}}$ in (35); $\boldsymbol{b}^{\text{GAS}}$ is a vector composed of the equivalent energy storage parameters to be calculated, that is

$$\boldsymbol{b}^{\text{GAS}} := \left[ (\overline{\boldsymbol{p}}^{\text{GAS}})^\top \;\; -(\underline{\boldsymbol{p}}^{\text{GAS}})^\top \;\; (\overline{\boldsymbol{e}}^{\text{GAS}})^\top \;\; -(\underline{\boldsymbol{e}}^{\text{GAS}})^\top \right]^\top \quad (37)$$

where $\overline{\boldsymbol{p}}^{\text{GAS}}$, $\underline{\boldsymbol{p}}^{\text{GAS}}$, $\overline{\boldsymbol{e}}^{\text{GAS}}$, and $\underline{\boldsymbol{e}}^{\text{GAS}}$ are vectors composed of the parameters $\overline{p}_t^{\text{GAS}}$, $\underline{p}_t^{\text{GAS}}$, $\overline{e}_t^{\text{GAS}}$, and $\underline{e}_t^{\text{GAS}}$ of all time slots, respectively.

To fully leverage the flexibility of the gas network, we try to find a group of parameters $\boldsymbol{b}^{\text{GAS}}$ to describe the equivalent energy storage model $\mathcal{B}^{\text{GAS}}$ as large as possible. To calculate the parameters $\boldsymbol{b}^{\text{GAS}}$ in (37), we use the high-dimensional polyhedron projection and bounds shrinking method proposed in our previous work [26], which is a two-stage method.

Denote the value of parameters in the $k$-th iteration as $\boldsymbol{b}_{(k)}^{\text{GAS}}$. Denote the equivalent energy storage model determined by parameter $\boldsymbol{b}_{(k)}^{\text{GAS}}$ in the $k$-th iteration as $\mathcal{B}_{(k)}^{\text{GAS}}$. In the first stage, we find out the scenarios that are not feasible in the current equivalent energy storage model $\mathcal{B}_{(k)}^{\text{GAS}}$. In the second stage, the parameters $\boldsymbol{b}_{(k)}^{\text{GAS}}$ are further modified so that these detected infeasible scenarios will not include in $\mathcal{B}_{(k+1)}^{\text{GAS}}$.

At first, the initial value $\boldsymbol{b}_{(0)}^{\text{GAS}}$ of parameters in this iteration process can be obtained by solving the following optimization problems:

$$\overline{p}_{t,(0)}^{\text{GAS}} = \left\{ \max_{\boldsymbol{z}_T^{\text{CTRL}}, \boldsymbol{p}^{\text{GAS}}} p_t^{\text{GAS}} \,\middle|\, (\boldsymbol{z}_T^{\text{CTRL}}, \boldsymbol{p}^{\text{GAS}}) \in \Omega^{\text{GAS}} \right\} \quad (38a)$$

$$\underline{p}_{t,(0)}^{\text{GAS}} = \left\{ \min_{\boldsymbol{z}_T^{\text{CTRL}}, \boldsymbol{p}^{\text{GAS}}} p_t^{\text{GAS}} \,\middle|\, (\boldsymbol{z}_T^{\text{CTRL}}, \boldsymbol{p}^{\text{GAS}}) \in \Omega^{\text{GAS}} \right\} \quad (38b)$$

$$\overline{e}_{t,(0)}^{\text{GAS}} = \left\{ \max_{\boldsymbol{z}_T^{\text{CTRL}}, \boldsymbol{p}^{\text{GAS}}} \sum_{\tau=0}^{t} p_\tau^{\text{GAS}} \cdot \Delta t \,\middle|\, (\boldsymbol{z}_T^{\text{CTRL}}, \boldsymbol{p}^{\text{GAS}}) \in \Omega^{\text{GAS}} \right\} \quad (38c)$$

$$\underline{e}_{t,(0)}^{\text{GAS}} = \left\{ \min_{\boldsymbol{z}_T^{\text{CTRL}}, \boldsymbol{p}^{\text{GAS}}} \sum_{\tau=0}^{t} p_\tau^{\text{GAS}} \cdot \Delta t \,\middle|\, (\boldsymbol{z}_T^{\text{CTRL}}, \boldsymbol{p}^{\text{GAS}}) \in \Omega^{\text{GAS}} \right\} \quad (38d)$$

Then, in the first stage, we find out the infeasible scenario by solving the following Stackelberg game [23], [26] problem:

$$f_{(k)}(\boldsymbol{b}_{(k)}^{\text{GAS}}) = \min_{\boldsymbol{p}^{\text{GAS}}} \max_{\boldsymbol{z}_T^{\text{CTRL}}} 0 \quad (39a)$$

$$s.t. \quad \boldsymbol{A}^{\text{GAS}} \boldsymbol{p}^{\text{GAS}} \leq \boldsymbol{b}_{(k)}^{\text{GAS}} \quad (39b)$$

$$\boldsymbol{W}_T \boldsymbol{z}_T^{\text{CTRL}} \leq \boldsymbol{w}_T \quad (39c)$$

$$\boldsymbol{p}^{\text{GAS}} = \boldsymbol{L} \boldsymbol{z}_T^{\text{CTRL}} \quad (39d)$$

In this Stackelberg game problem, the leader is any possible dispatch scenario described by $\boldsymbol{p}^{\text{GAS}}$ with the flexibility of equivalent model $\mathcal{B}_{(k)}^{\text{GAS}}$, and the follower is the control action that guarantee all the constraints (31) and (33) hold. If there is no control actions in $\Omega^{\text{GAS}}$ can realize a specific dispatch scenario in $\mathcal{B}_{(k)}^{\text{GAS}}$, it means this scenario is infeasible, and the infeasible point is denoted as $\boldsymbol{p}_{(k)}^{\text{GAS}}$. Then, the optimal value of $f_{(k)}^*$ is $-\infty$, because the maximum problem in (39) is infeasible. On the contrary, if all the scenarios inside the $\mathcal{B}_{(k)}^{\text{GAS}}$ is feasible for the gas network, the optimal value $f_{(k)}^*$ equals 0.

To solve the Stackelberg game (39), the inner problem can be transferred into its dual minimize problem. Then it will become a bilinear optimization problem.

$$\min_{\boldsymbol{p}^{\text{GAS}}, \boldsymbol{v}, \boldsymbol{\mu}} \quad \boldsymbol{v}^\top \boldsymbol{w}_T + \boldsymbol{\mu}^\top \boldsymbol{p}^{\text{GAS}} \tag{40a}$$

$$s.t. \quad \boldsymbol{A}^{\text{GAS}} \boldsymbol{p}^{\text{GAS}} \leq \boldsymbol{b}_{(k)}^{\text{GAS}} \tag{40b}$$

$$\boldsymbol{v}^\top \boldsymbol{W}_T + \boldsymbol{\mu}^\top \boldsymbol{L} = \boldsymbol{0} \tag{40c}$$

$$\boldsymbol{v} \geq \boldsymbol{0} \tag{40d}$$

where $\boldsymbol{v}$ and $\boldsymbol{\mu}$ are the dual variables of constraints (39c) and (39d), respectively.

Subsequently, we can use the KKT condition of (40b) and introduce auxiliary binary variables to transform (40) into the following MILP problem:

$$\min_{\boldsymbol{p}^{\text{G}}, \boldsymbol{v}, \boldsymbol{\mu}, \boldsymbol{\zeta}, \boldsymbol{s}} \quad \boldsymbol{v}^\top \boldsymbol{w}_T - \boldsymbol{\zeta}^\top \boldsymbol{b}_{(k)}^{\text{GAS}} \tag{41a}$$

$$s.t. \quad \boldsymbol{\zeta}^\top \boldsymbol{A}^{\text{GAS}} + \boldsymbol{\mu}^\top = \boldsymbol{0} \tag{41b}$$

$$\boldsymbol{v}^\top \boldsymbol{W}_T + \boldsymbol{\mu}^\top \boldsymbol{L} = \boldsymbol{0} \tag{41c}$$

$$\boldsymbol{v} \geq \boldsymbol{0} \tag{41d}$$

$$0 \leq \boldsymbol{b}_{(k)}^{\text{GAS}} - \boldsymbol{A}^{\text{GAS}} \boldsymbol{p}^{\text{GAS}} \leq M(\boldsymbol{1} - \boldsymbol{s}) \tag{41e}$$

$$0 \leq \boldsymbol{\zeta} \leq M\boldsymbol{s} \tag{41f}$$

where $\boldsymbol{\zeta}$ denotes the dual variable of (40b). $\boldsymbol{s}$ is vector composed of binary variables and has the same dimension as $\boldsymbol{b}_{(k)}^{\text{GAS}}$. $\boldsymbol{1}$ denotes an all-one vector; $M$ is a sufficiently large constant. (41e)-(41f) are the inequities equivalent to the complementary slackness condition of the KKT condition of (40b).

Since the optimal value is always obtained in the extreme point of the feasible region, $\boldsymbol{p}_{(k)}^{\text{GAS}}$ must be an extreme point of $\mathcal{B}_{(k)}^{\text{GAS}}$. Then we can find out the active constraints and denote the index set of them as $\mathcal{P}_{(k)}$, that is

$$\left(\boldsymbol{A}^{\text{GAS}}\right)_i \boldsymbol{p}_{(k)}^{\text{GAS}} = \left(\boldsymbol{b}_{(k)}^{\text{GAS}}\right)_i, \forall i \in \mathcal{P}_{(k)} \tag{42}$$

where $(\cdot)_i$ denotes the $i$-th row of a matrix or the $i$-th element of a vector.

Then we can obtain a feasible point in the control feasible region that is closest to $\boldsymbol{p}_{(k)}^{\text{GAS}}$ by solving the optimization problem (43). This feasible point must be on the boundary of the feasible region and denoted as $\boldsymbol{p}_{(k)}^{\text{bd}}$.

$$\boldsymbol{p}_{(k)}^{\text{bd}} = \arg\min_{\boldsymbol{p}^{\text{GAS}}} \left\| \boldsymbol{p}_{(k)}^{\text{GAS}} - \boldsymbol{p}^{\text{GAS}} \right\|_2^2 \tag{43a}$$

$$s.t. \quad \left(\boldsymbol{z}_T^{\text{CTRL}}, \boldsymbol{p}^{\text{GAS}}\right) \in \Omega^{\text{GAS}} \tag{43b}$$

In the second stage, the flexibility region of equivalent model $\mathcal{B}_{(k)}^{\text{GAS}}$ will shrink by adjusting the parameters $\boldsymbol{b}_{(k)}^{\text{GAS}}$ related to the active constraints in (42). A new group of parameters $\boldsymbol{b}_{(k+1)}^{\text{GAS}}$ will be calculated according to (44) as follows:

$$\boldsymbol{b}_{(k+1)}^{\text{GAS}} = \arg\max_{\boldsymbol{b}^{\text{GAS}}, \boldsymbol{w}} \quad \boldsymbol{1}_{n_b}^\top \boldsymbol{b}^{\text{GAS}} \tag{44a}$$

$$s.t. \quad \boldsymbol{b}^{\text{GAS}} \leq \boldsymbol{b}_{(k)}^{\text{GAS}} \tag{44b}$$

$$\boldsymbol{1}_{n_k}^\top \boldsymbol{w} \geq T, \quad \boldsymbol{w} \in \{0,1\}^{n_k} \tag{44c}$$

For all $i \in \mathcal{P}_{(k)}$

$$\left(\boldsymbol{A}^{\text{GAS}}\right)_i \boldsymbol{p}_{(k)}^{\text{bd}} \geq \left(\boldsymbol{b}^{\text{GAS}}\right)_i - M(1 - w_j) \tag{44d}$$

where $n_k := \text{card}(\mathcal{P}_{(k)})$; $n_b$ denotes the length of vector $\boldsymbol{b}^{\text{GAS}}$; $\boldsymbol{w}$ is a $n_k$-dimensional binary variable vector.

Then, this new group of parameters $\boldsymbol{b}_{(k+1)}^{\text{GAS}}$ will go back to the first stage for further control feasibility check. Finally, the equivalent energy storage model $\mathcal{B}^{\text{GAS}}$ can be obtained.

The pseudocode of this two-stage iteration process is described in Algorithm 1.

| **Algorithm 1.** The high-dimensional polyhedron projection and bounds shrinking algorithm |
|---|
| 1: Initialize $k=0$, calculate (38) to obtain $\boldsymbol{b}_{(0)}^{\text{GAS}}$ |
| 2: Solve the Stackelberg game problem (39). Obtain the optimal objective value $f_{(k)}^*$ and the potential infeasible extreme point $\boldsymbol{p}_{(k)}^{\text{GAS}}$. |
| 3: If $f_{(k)}^* = 0$, output $\boldsymbol{b}_{(k)}^{\text{GAS}}$ as the final results and exit the algorithm, otherwise go to step 4. |
| 4: Solve (43) and obtain the boundary point $\boldsymbol{p}_{(k)}^{\text{bd}}$ in $\Omega^{\text{GAS}}$ that is closest to $\boldsymbol{b}_{(k)}^{\text{GAS}}$. |
| 5: Update the value of $\boldsymbol{b}_{(k)}^{\text{GAS}}$ to $\boldsymbol{b}_{(k+1)}^{\text{GAS}}$ based on (44). |
| 6: $k \leftarrow k+1$ and go back to step 2. |

*D. The integrated equivalent model of the gas network*

By integrating the power coupling region of devices $\mathcal{E}_t$ of all time slots in Section 3.1 and the energy storage flexibility region $\mathcal{B}^{\text{GAS}}$ in Section 3.2, the flexibility model of gas network can be finally obtained.

Define the variable matrix $\boldsymbol{P}^{\text{DEV}}$ that composed of the active power of GTs and P2Gs of all time slots. The row number of $\boldsymbol{P}^{\text{DEV}}$ represents the index of device, and the column number represents the index of time slots. According to the definitions of $\boldsymbol{p}_t^{\text{DEV}}$ and $\boldsymbol{p}^{\text{GAS}}$ in (20) and (32), they can be expressed by $\boldsymbol{P}^{\text{DEV}}$ in a linear form:

$$\boldsymbol{p}_t^{\text{DEV}} = \boldsymbol{P}^{\text{DEV}} \boldsymbol{\delta}_t, \forall t \in \mathcal{T} \quad (45a)$$

$$\boldsymbol{p}^{\text{GAS}} = \left(\boldsymbol{\eta}^\top \boldsymbol{P}^{\text{DEV}}\right)^\top \quad (45b)$$

where $\boldsymbol{\delta}_t$ is a constant vector, whose $t$-th element is 1, and others are all 0; vector $\boldsymbol{\eta}$ is defined in (32).

Finally, the integrated flexibility model of gas network is as follows:

$$\mathcal{F}^{\text{GAS}} := \left\{ \boldsymbol{P}^{\text{DEV}} \left| \begin{array}{l} \boldsymbol{P}^{\text{DEV}} \boldsymbol{\delta}_t \in \mathcal{E}_t, \forall t \in \mathcal{T} \\ \left(\boldsymbol{\eta}^\top \boldsymbol{P}^{\text{DEV}}\right)^\top \in \mathcal{B}^{\text{GAS}} \end{array} \right. \right\} \quad (46)$$

## V. NUMERICAL TEST CASES

### A. Case setup

The case studies are carried on an integrated system of the IEEE RTS96 One Area 24 bus power network [27] and the GasLib-134 [28] Greek gas network with some modifications. There are 4 GTs, 4 P2Gs, 5 compressors and 2 GWs in the gas network. All the parameters of units and the topology of gas network are listed in the supplemental file [29].

All the numerical cases are conducted on a laptop with Intel Core i7-1165G7 CPU and 16 GB RAM. The programming are implemented in MATLAB R2020b, with GUROBI [30] as the solver and YALMIP [31] as the model tool of optimization problems.

### B. Result of the converters' power coupling region

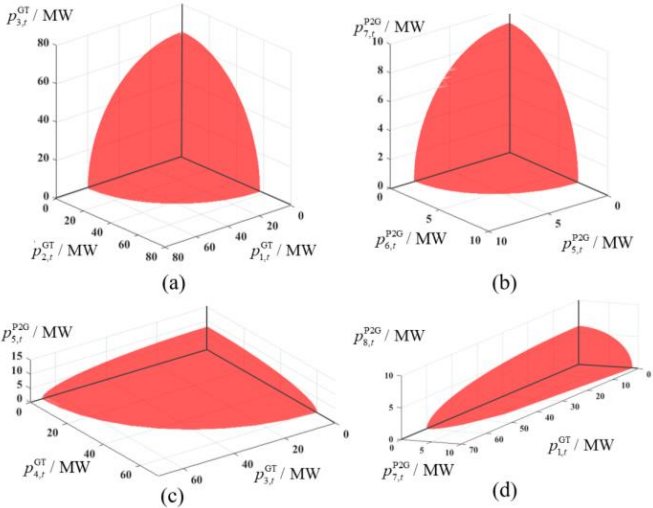

Fig. 4. Projection of high-dimensional quadrant ellipsoid in the 3-dimensional space $(t=12:00)$.

The power coupling region among different energy exchange interfaces $\mathcal{E}_t$ of each time slot is calculated. Since there are eight GTs and P2Gs in total, the region is an eight-dimensional quadrant ellipsoid. However, we cannot draw more than three-dimensional space images, but we can visualize it by selecting three variables at a time to draw the projection of this eight-dimensional ellipsoid to the three-dimensional space. Fig. 4 shows coupling region among different energy exchange interfaces at 12:00. The region is presented in the form of a group of three-dimensional ellipsoids, which is the projection results from eight-dimensional space. In each graph, the constraint relationship among the selected three different variables is visualized in a three-dimensional quadrant ellipsoid form.

To demonstrate the approximating accuracy resulting from the aforementioned inscribed quadrant ellipsoid model, we generated 1000 operational scenarios of the gas network using Monte Carlo simulation. All generated scenarios are technically feasible, residing within the yellow region depicted in Fig. 3. Due to the approximation error of the inscribed model, some scenarios are encompassed within the quadrant ellipsoid, corresponding to the orange region in Fig. 3. The proportion of scenarios within the inscribed quadrant ellipsoid represents the coverage rate of the approximation model, reflecting the precision of the inscribed quadrant ellipsoid. The coverage rate of all time slots are shown as follows:

TABLE I
THE COVERAGE RATE OF ALL TIME SLOTS

| Maximum | Average | Minimum |
|---------|---------|---------|
| 100.0%  | 98.7%   | 96.9%   |

The results indicate that the inscribed quadrant ellipsoid model exhibits a higher coverage rate. Hence, the quadrant ellipsoid effectively encompasses the majority of operational scenarios.

### C. Result of the energy storage flexibility region

The energy storage flexibility region $\mathcal{B}^{\text{GAS}}$ is calculated based on the high-dimensional polyhedron projection and bounds shrinking algorithm, which describes the time-varying capacity of the equivalent energy storage model. The upper and lower bounds of aggregated power of gas network is shown in Fig. 5, corresponding to the parameters $\overline{p}_t^{\text{GAS}}$, $\underline{p}_t^{\text{GAS}}$ in (35a). The upper and lower energy capacity bounds of gas network is shown in Fig. 6, corresponding to the parameters $\overline{e}_t^{\text{GAS}}$, $\underline{e}_t^{\text{GAS}}$ in (35b).

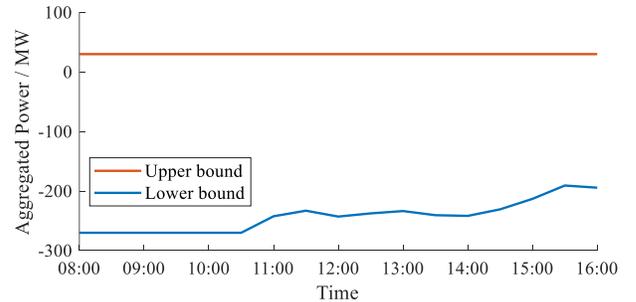

Fig. 5. The upper and lower aggregated power bounds of gas network.

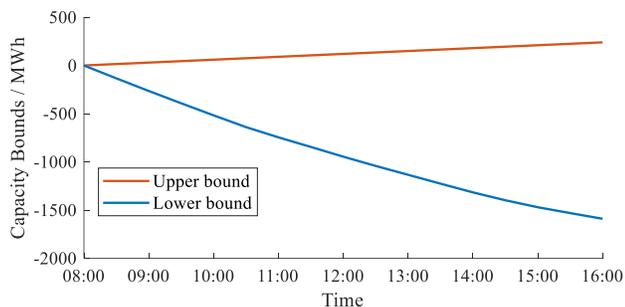

Fig. 6. The upper and lower energy capacity bounds of gas network.

In the equivalent energy storage model, the GTs act as discharge devices, while the P2Gs serve as charging devices. With the passage of time, both the maximum charging and discharging energy bounds gradually increase, causing a gradual expansion of the upper and lower energy bounds of the equivalent energy storage model. The upper bounds of power and energy are formed by the P2Gs. Since the capacity of P2Gs are relative small, their maximum active powers are seldom limited. Therefore, the upper bounds are usually straight lines. On the contrary, the lower bounds are formed by the GTs and their maximum active powers may be limited by the operation constraints of gas network, such as the constraints of gas pressure.

By integrating the former results of $\mathcal{E}_t$ and $\mathcal{B}^{\text{GAS}}$, the integrated flexibility model of gas network $\mathcal{F}^{\text{GAS}}$ can be finally obtained.

### D. Comparison of calculation efficiency and dispatch error

The calculated equivalent model of gas network can be easily incorporated into the optimal dispatch problem of the power system and reduce the computational burden. In this numerical case, we compare the calculation time and dispatch error of different joint dispatch methods.

In this case, we randomly generate 100 scenarios with different load curves. Subsequently, for each scenario, we use the detailed gas network model (as the benchmark), the static gas network equivalent model proposed in [17], and our gas network equivalent model for joint optimal dispatch of electricity-gas systems, respectively. For the benchmark method, the detailed model of the gas network is used, which is solved by the finite difference method (FDM). It can be considered as the accurate results. The comparisons of calculation time and dispatch error are listed in TABLE II.

TABLE II
COMPARISONS OF CALCULATION TIME AND DISPATCH ERROR

| Comparison items | Calculation time / s | | | Percentage of dispatch error | |
|---|---|---|---|---|---|
| | FDM | Static model | Our model | Static model | Our model |
| Maximum | 94.71 | 0.71 | 0.83 | 4.45% | 0.89% |
| Average | 75.38 | 0.35 | 0.42 | 2.08% | 0.25% |
| Minimum | 67.15 | 0.19 | 0.23 | 0.00% | 0.00% |

The results in TABLE II show that the proposed equivalent energy storage model can effectively reduce the calculation time of the optimal dispatch problem with high accuracy.

Furthermore, our gas network equivalent model can obtain much higher accurate results than the static equivalent one. This is because our equivalent model is calculated using the gas network state-space model, rather than the steady state equation. Therefore, it can fully exploit the flexibility of gas networks. In addition, the detailed parameters of the gas network are not explicitly used in the proposed calculation process, so the information privacy of gas networks is preserved.

Subsequently, to verify the feasibility of the resulted control strategies of equivalent models. We use Monte Carlo simulations to randomly generate 10,000 scenarios which are solved using the static equivalent model in [17] and our model, respectively. Each sample represents the dispatch plan of all the units for each time slot. We check whether the dispatch plan of all units meets the gas network operation constraints. The proportion of infeasible scenarios is used as an indicator to measure the feasibility of the resulted control strategies of different models, and the results are shown in TABLE III.

TABLE III
PROPORTION OF INFEASIBLE SCENARIOS

| Methods | Static model | Our model |
|---|---|---|
| Proportion of infeasible scenarios | 8.37% | 0.00% |

The results show that our equivalent model can guarantee the control strategies feasible in all the scenarios, while the static model may cause infeasible results. This is because our model incorporates the temporal coupling constraints as well as the transient process of gas networks.

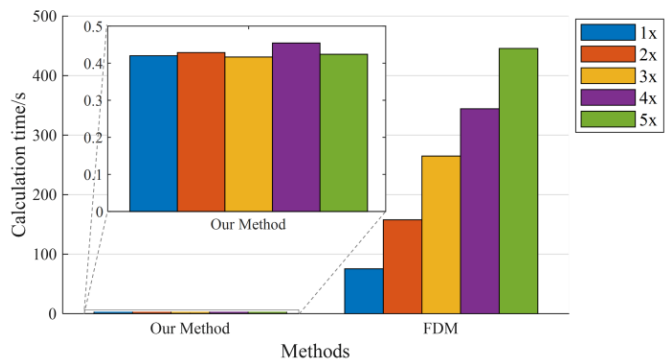

Fig. 7. Comparison of calculation time for different scale systems using two different methods.

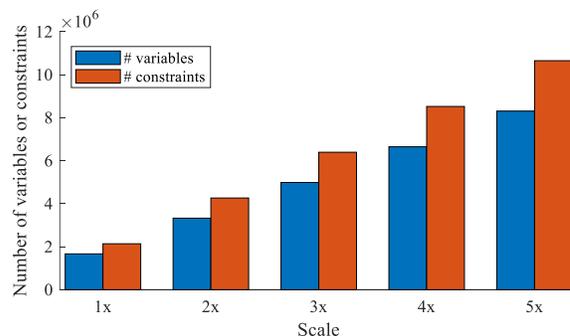

Fig. 8. Number of variables and constraints with FDM as the scale of gas network expands.

To illustrate the scalability of our method, we scale up the gas networks and test the calculation time of each scale system. "1x" represents the original scale of the gas network, and "2x"-"5x" represent the gas networks with a scale of 2-5 times, respectively. The results are shown in Fig. 7 and Fig. 8.

According to Fig. 7, the calculation time of FDM increases fast with the expansion of the system scale, while the calculation time of our method is almost unchanged. Based on Fig. 8, it can be observed that the number of variables and constraints associated with FDM are more than millions, and they increase linearly as the scale expands. However, the complexity of our method remains unaffected by the scale of the gas network. The number of variables and constraints remains constants at 5,760 and 7,104, respectively.

## VI. Conclusions

This work presented the detailed process for evaluating the equivalent energy storage model of the gas network. The state-space model of gas networks is reduced to a multi-port energy storage model with time-varying capacity, which implicitly incorporates the dynamic gas state transformation process and all operational constraints. This equivalent model can make full use of the flexibility of gas networks with guaranteed feasibility. From the perspective of the power system, the proposed method transforms the complex state-space model of the gas network into a simplified energy storage model. The reduced equivalent model can be easily incorporated into the optimal dispatch problem of the power system and the information privacy of gas networks is preserved.

Numerical tests demonstrate the superior efficacy of the equivalent energy storage model in mitigating the computational burden associated with the coordinated optimal dispatch of electricity-gas systems, while maintaining high accuracy. In the joint optimal dispatch problem, our method yields a remarkable reduction in computation time, decreasing it from an average of 75s with the finite difference method to a mere 0.42s, thereby achieving a significant decrease of over two orders of magnitude. Moreover, our method exhibits an average dispatch error of only 0.25%. Furthermore, in comparison to existing static models that would yield an infeasible scenario rate of 8.37%, our model ensures the feasibility of all resulting control strategies.